\documentclass[prd,showpacs,preprintnumbers,amsmath,amssymb,twocolumn,superscriptaddress,notitlepage]{revtex4-2}

\usepackage{cases}
\usepackage{amsmath}
\usepackage{amssymb}
\usepackage{amsfonts}
\usepackage{amssymb}
\usepackage{dcolumn}
\usepackage{bm}
\usepackage{epsfig}
\usepackage{bbm}
\usepackage{graphicx}
\usepackage{xcolor}
\usepackage{array}
\usepackage{subfigure}
\usepackage{hyperref}
\usepackage{mathrsfs}
\usepackage{verbatim}
\usepackage{float}
\usepackage{epstopdf}
\usepackage{color}
\usepackage{orcidlink}

\newcommand{\be}{\begin{equation}}
\newcommand{\ee}{\end{equation}}
\newcommand{\ba}{\begin{eqnarray}}
\newcommand{\ea}{\end{eqnarray}}

\newcommand{\gsim}{\mathrel{\hbox{\rlap{\lower.55ex \hbox {$\sim$}}
			\kern-.3em \raise.4ex \hbox{$>$}}}}
\newcommand{\lsim}{\mathrel{\hbox{\rlap{\lower.55ex \hbox {$\sim$}}
			\kern-.3em \raise.4ex \hbox{$<$}}}}

\usepackage{ulem,fancyvrb}
\usepackage{xcolor}

\hypersetup{colorlinks=true,
	breaklinks=true,
	pdfstartview=Fit,
	linkcolor=blue,
	citecolor=blue,
	urlcolor=blue}

\begin{document}
\title{
Testing the Fifth Force on Lepton Spins through Neutrino Oscillations}

\author{Rundong Fang${}^\ast$} 
\thanks{These authors contributed equally to this work.}
\affiliation{School of Physics, Beihang University, Beijing 100083, China}

\author{Ji-Heng Guo${}^\ast$}
\thanks{These authors contributed equally to this work.}
\affiliation{School of Physics, Beihang University, Beijing 100083, China}
 
\author{Jia Liu \orcidlink{0000-0001-7386-0253}}
\email[Contact author: ]{jialiu@pku.edu.cn}
\affiliation{School of Physics and State Key Laboratory of Nuclear Physics and Technology, Peking University, Beijing 100871, China}
\affiliation{Center for High Energy Physics, Peking University, Beijing 100871, China}

\author{Xiao-Ping Wang \orcidlink{0000-0002-2258-7741} }
\email[Contact author: ]{hcwangxiaoping@buaa.edu.cn}
\affiliation{School of Physics, Beihang University, Beijing 100083, China}
\affiliation{Beijing Key Laboratory of Advanced Nuclear Materials and Physics, Beihang University,
Beijing 100191, China}

\begin{abstract}
We investigate a fifth force mediated by a light vector boson that couples to lepton spins, characterized by axial-vector couplings to leptons and vector couplings to nucleons. This interaction generates a potential proportional to the inner product of the lepton spin vector and the nucleon-lepton relative velocity vector, a feature extensively explored with precision spin sensors. Employing weak symmetry, we show that left-handed charged lepton couplings  naturally extend to left-handed neutrinos, enabling this fifth force to influence neutrino oscillations. For electron-nucleon couplings, we find that solar and reactor neutrino experiments provide comparable constraints to those from spin sensors and surpass them in the short-range fifth force region. For muon-nucleon couplings, neutrino oscillation experiments exclude the fifth force as a viable explanation for the muon $ g-2 $ anomaly in the context of a vector mediator, tightening the bounds by two orders of magnitude in coupling strength by solar and atmospheric neutrino data. Our results highlight the critical role of neutrino oscillations in probing fifth forces acting across all three generations of lepton spins.
\end{abstract}

\maketitle

\section{Introduction}

Physics beyond the Standard Model (SM) often predicts novel interactions that go beyond the four known fundamental forces, commonly referred to as the ``fifth force.” Many of these interactions involve the spin of particles~\cite{Cong:2024qly}, and experimental sensitivity can be achieved through spin sensor experiments. These experiments make it possible to measure the fifth force by detecting interactions with either electron spin~\cite{Ni:1999di, Chu:2015tha, Crescini:2020ykp, Vorobev:1988ug, Huang:2024esg} or nucleon spin~\cite{Venema:1992zz, Glenday:2008zz, Vasilakis:2008yn, Su:2021dou, Tullney:2013wqa, Wei:2022mra, Zhang:2023qmu}.

If the fifth force is mediated by a very light boson, it could operate over long distances. This mediator, typically a scalar or vector particle, couples to nucleons via scalar or vector interactions, and to leptons via pseudoscalar or axial-vector (AV) interactions. As a result, large astronomical bodies, such as the Sun, Earth, and Moon, which contain vast numbers of particles, can serve as sources for this force. 
In the non-relativistic limit, they give rise to a potential that is related to charged lepton or nucleon spins~\cite{Dobrescu:2006au, Fadeev:2018rfl}, which can be constrained through precision spin sensor experiments.
This opens up the possibility for precision spin sensors to set stringent constraints on the fifth force models~\cite{Heckel:2008hw, Yan:2014ruk, Wei:2022mra, Wu:2023ypz, Clayburn:2023eys}.

For a fifth force coupling to lepton spins, weak symmetry suggests that left-handed lepton couplings may extend to left-handed neutrinos. As a result, the same fifth force could influence the evolution of neutrino states as they propagate, allowing neutrino oscillation experiments to serve as tests of this force.

For neutrino oscillations, most prior studies have focused on neutrino non-standard interactions (NSI)~\cite{Wolfenstein:1977ue, Farzan:2017xzy, 10.21468/SciPostPhysProc.2.001}, assuming a heavy mediator that generates a potential proportional to local density. A number of works, however, have explored long-range interactions in Refs.~\cite{Gonzalez-Garcia:2006vic, Bandyopadhyay:2006uh, Gonzalez-Garcia:2008jtw, Samanta:2010zh,
Wise:2018rnb, Babu:2019iml,
Smirnov:2019cae, 
Chauhan:2024qew, Ansarifard:2024zxm,
Coloma:2020gfv, Agarwalla:2024ylc, Mishra:2024riq} for the mediator mass is less than $\sim 10^{-12}$ eV and Refs.~\cite{Denton:2020uda, Esteban:2021ozz} for larger mediator mass. 
These studies generally examine vector-vector (V-V) or scalar-scalar (S-S) interactions, where the primary contributions are spin-independent. 

In this work, we explore a fifth force mediated by a light vector boson that couples to lepton spins, extending beyond the scalar mediator studied in our previous work~\cite{Fang:2024syu}. The mediator is characterized by vector and axial-vector (V-AV) couplings to nucleons and leptons, respectively. We demonstrate that this type of fifth force can be tested through neutrino oscillation experiments. For electron-nucleon couplings, both solar and reactor neutrino experiments impose stringent and complementary constraints compared to spin sensor experiments and surpass them in the short-range fifth force region. For muon-nucleon couplings, our analysis shows that neutrino oscillations rule out the fifth force explanation for the muon \( g-2 \) anomaly~\cite{Agrawal:2022wjm, Davoudiasl:2022gdg} with a vector mediator, tightening the coupling strength constraints by two orders of magnitude. Therefore, neutrino oscillations provide a crucial test for the existence of a fifth force acting on lepton spins.

\section{Model setup}

We consider a $\rm U(1)'$ vector gauge boson $A'$ couples to leptons ($L$) and nucleons ($N$) through AV and vector couplings, respectively. The general form of its interaction Lagrangian is given by:
\begin{align}
\mathcal{L}_{\rm int} = &A'_{\mu}
\left( \boldsymbol{g}_L^{ij} \bar{L}^{i} \gamma^{\mu} L^{j} + \boldsymbol{g}_R^{ij} \bar{e}_R^{i} \gamma^{\mu} e_R^{j} \right)
+ g_V^N A'_\mu \bar{N} \gamma^\mu N,
\label{eq:5th force_lagrangian_ava}
\end{align}
where $i$ and $j$ are flavor indices and we assume the nucleon couplings are identical for neutrons and protons.
Since we require $A'$ coupling to leptons in an AV coupling, therefore, we have
\begin{align}
\mathcal{L}_{\rm int} & =  \boldsymbol{g}_{A}^{ij} A'_{\mu}
\left( - \bar{\nu}_L^i \gamma^{\mu} \nu_L^j +  \bar{e}^{i} {\gamma}^{\mu} \gamma^5 e^{j}  \right)\nonumber \\
&+ g_V^N A'_\mu \left( \bar{p} \gamma^\mu p + \bar{n} \gamma^\mu n \right),
\label{eq:5th force_lagrangian_ava-2}
\end{align}
where $\boldsymbol{g}_A = -\boldsymbol{g}_L =  \boldsymbol{g}_R$
ensures that the coupling to leptons is purely axial-vector.
We assume a flavor-diagonal axial-vector coupling $\boldsymbol{g}_A^{ij} = a_i \delta^{ij}$. To ensure anomaly cancellation, the couplings should satisfy $\sum_i a_i = \sum_i a_i^3 = 0$~\cite{Ismail:2016tod}. This can be achieved via inter-generational cancellation, e.g., $a_\tau = -a_\mu$, $a_e = 0$, similar to the $L_\mu - L_\tau$ model.
However, to stay model-independent, we will calculate constraints for one generation at a time.

\section{fifth force potential} 

Considering light $A'$, the net baryon charge can behave as electromagnetic (EM) charge and generates the static $A'$ potential.
For a point source with net baryon charge of $N_n$, we can have the generated $A'$ in Yukawa potential as,
\be
A'_{0}(\vec{r}) = - g_V^N   \frac{N_n}{4 \pi r} e^{-m_{A'}r}, \quad 
\vec{A'}(\vec{r}) \simeq 0
\label{eq:vector_boson_field}
\ee
where $m_{A'}$ is the mass of $A'$, $r$ is the distance to the point source. 
The second equality because we assume the source does not have spin polarization. 
For an extended spherical source, e.g. Earth, with a local number density of nucleons $n(r)$, the static $A'$ potential becomes~\cite{Coloma:2020gfv}
\be
\small
A'_0(r)
= \frac{- g_V^N}{2 m_{A'}  r} \int_0^{R_\oplus} d r' n(r')r' \left( e^{- m_{A'} \left| r' - r\right|} -  e^{- m_{A'} \left( r + r' \right)}\right),
\label{eq:vector_field_full_calc}
\ee
where $R_\oplus$ is the radii of the source.
For very light $A'$, $m_{A'} \ll 1/R_\oplus$, one can recover the standard EM potential. For the heavy $A'$ case, $m_{A'} \gg 1/R_\oplus$, one can obtain~\cite{Coloma:2020gfv}  
\be
A'_0(r) \simeq - g_V^N \frac{n(r)}{m_{A'}^2} ,
\label{eq:approx_potential_2}
\ee
by localizing the integral range in Eq.~\eqref{eq:vector_field_full_calc} near $r$, e.g. $r' \in [r - 10 m_{A'}^{-1}, r + 10 m_{A'}^{-1}]$ and neglecting the small exponential suppression factor. Notably, this is similar to the MSW potential~\cite{Wolfenstein:1977ue, Mikheyev:1985zog} as expected.

\begin{figure*}[htb]
\begin{centering}
\includegraphics[width=0.32\textwidth]
{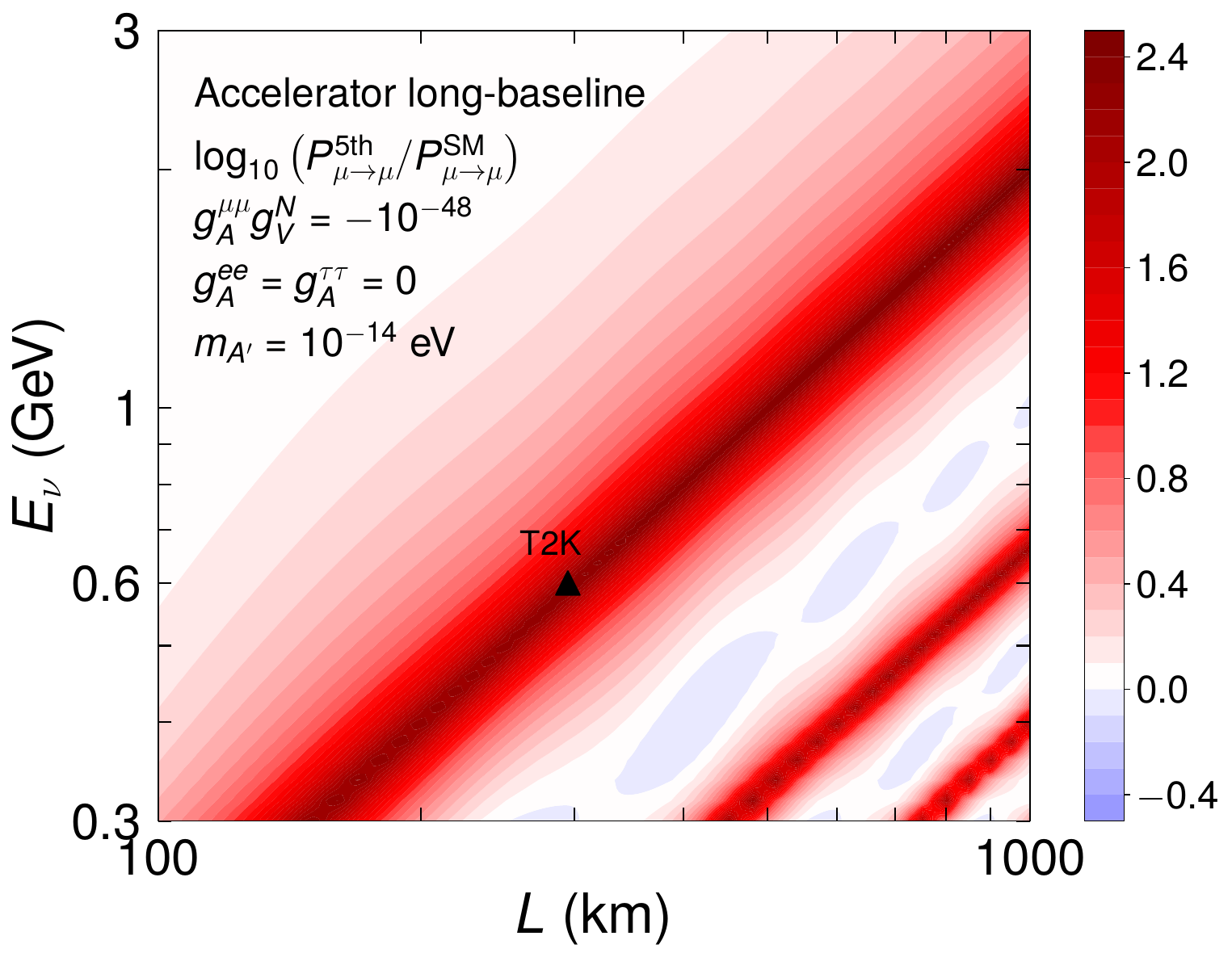}
\includegraphics[width=0.32\textwidth]{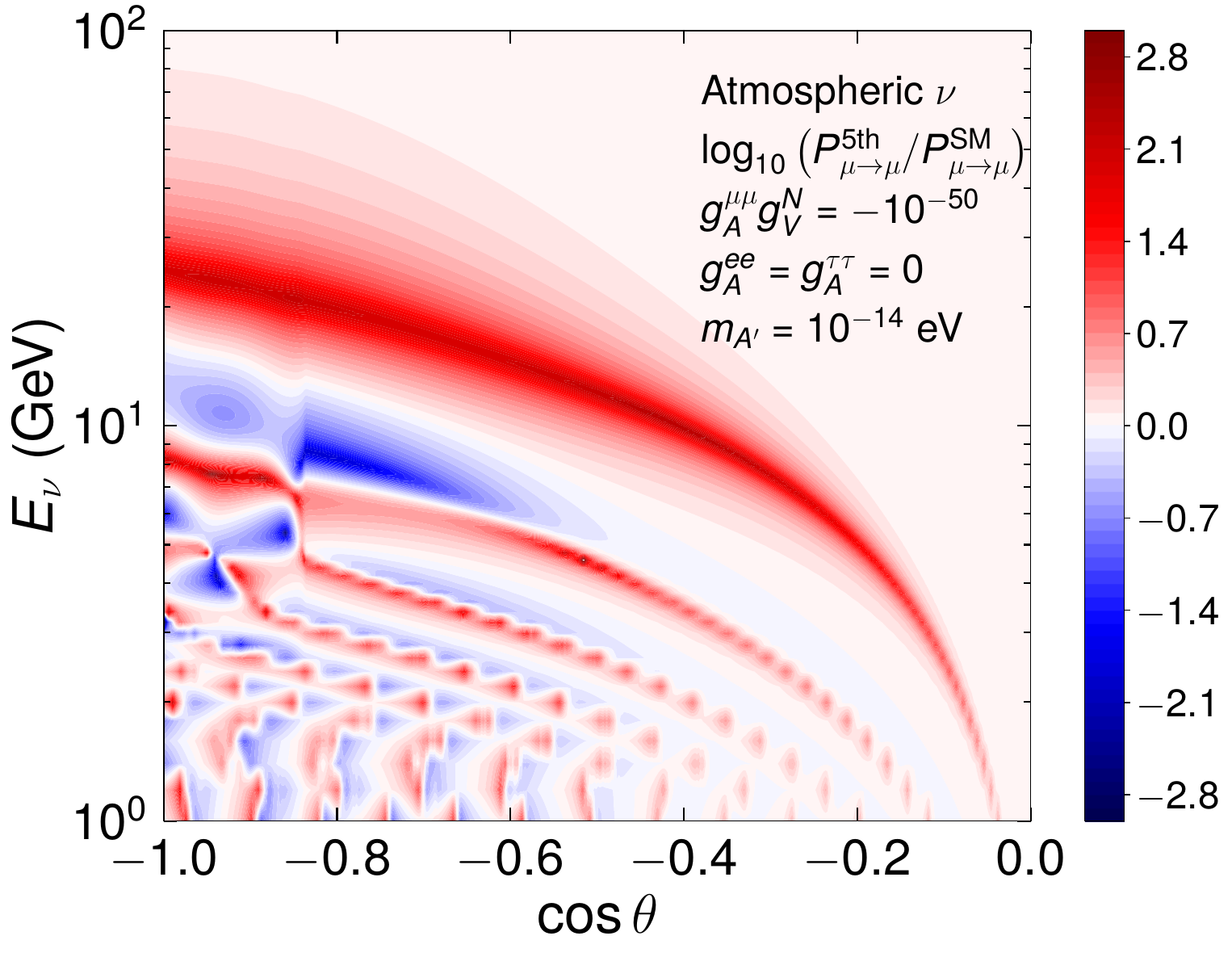}
\includegraphics[width=0.32\textwidth]
{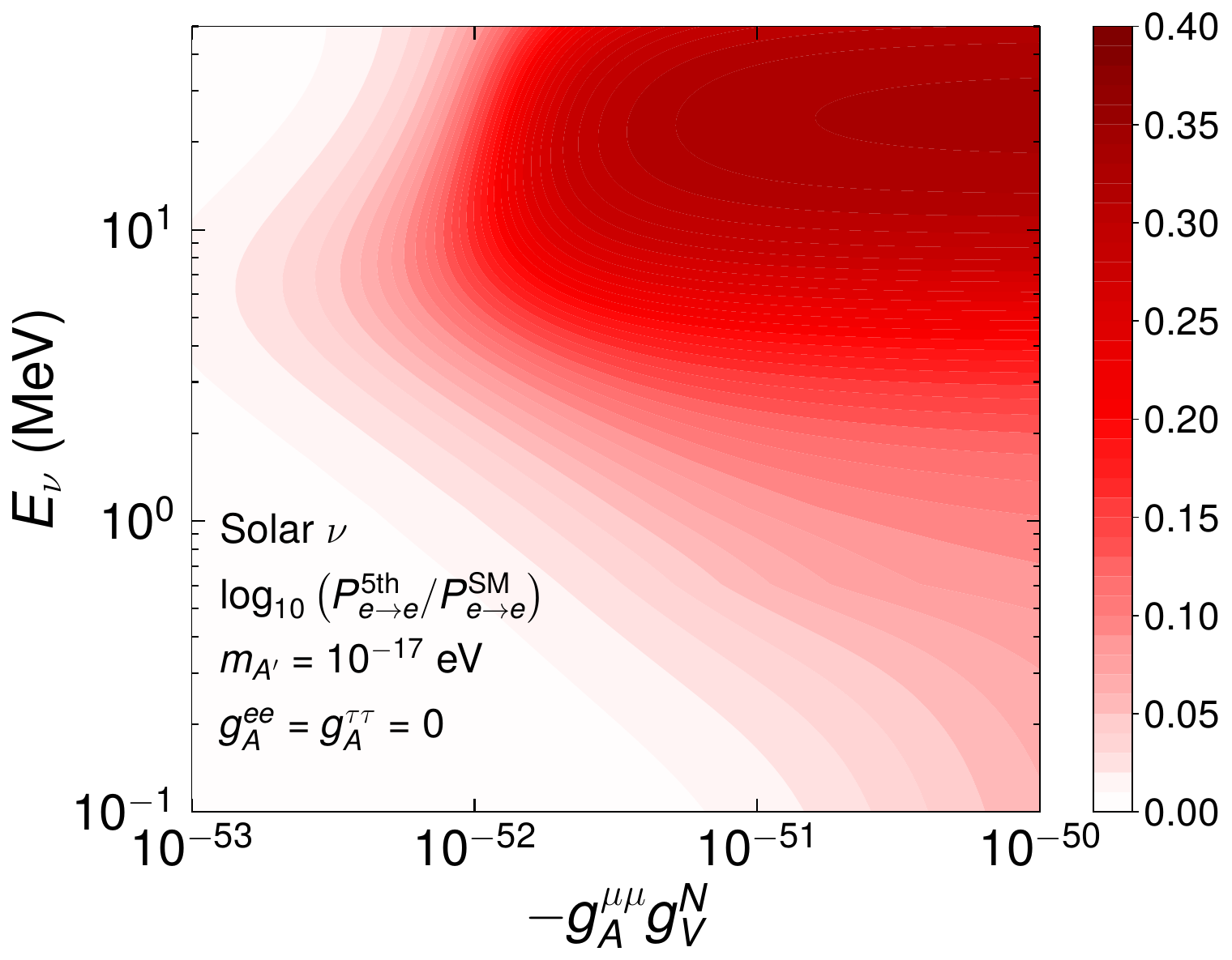}
\caption{The contour plots of ratios between the oscillation probability  with and without the fifth force. Left panel: the $\nu_\mu$ survival probability ratio as a function of neutrino energy $E_\nu$ and baseline length $L$, with the black triangle marking the typical neutrino energy and baseline for the T2K experiment. Central panel: the $\nu_\mu$ survival probability ratio as a function of $E_\nu$ and $\cos\theta$, relevant for IceCube experiments. Right panel: the solar neutrino disappearance probability ratio as a function of $E_\nu$ and the coupling $g_A g_V^N$. 
}
\label{fig:osc_fraction-all} 
\end{centering}
\end{figure*}

\section{Neutrino oscillations}

The neutrino oscillation probability can be straightforwardly calculated using the following Hamiltonian:
\begin{align}
\mathcal{H} 
=U \frac{M^2}{2 E_{\nu}}U^{\dagger} + V_{\rm MSW} + \boldsymbol{g}_{A} A'_0 \, ,
\label{eq:fullH}
\end{align}
where $U$ is the PMNS matrix~\cite{Pontecorvo:1957qd, Maki:1962mu} and $M$ is the neutrino mass matrix. The second term accounts for the MSW matter effect. The third term introduces the fifth force contribution, arising from $A'$ modifying the neutrino momentum as, $p_\mu \to p_\mu - \boldsymbol{g}_A A'_\mu$ \cite{Brdar:2017kbt, Huang:2018cwo, Fang:2024syu}. In the following analysis, we focus on the diagonal couplings $ \boldsymbol{g}_{A}^{ee} $ and $ \boldsymbol{g}_{A}^{\mu \mu} $, while generalizations to other couplings can be readily performed. The oscillation probabilities are computed through comprehensive numerical simulations with Hamiltonian \eqref{eq:fullH}. We analyze across various neutrino experiments, including accelerator and reactor long-baseline (LBL) experiments, reactor medium-baseline (MBL) experiments, atmospheric neutrino (ATM) experiments, and solar neutrino (Solar) experiments. 

For the reactor and accelerator experiments, we assume a constant interaction Hamiltonian $\Delta \mathcal{H}$ during propagation, taking Earth matter density as 2.7 ${\rm g}/{\rm cm}^{3}$ for MSW effects. For ATM neutrinos, we incorporate the variable Earth matter density along the neutrino’s path, segmenting the trajectory to calculate oscillation probability using PREM density data~\cite{Dziewonski:1981xy}. The production height of atmospheric neutrinos is fixed at 10 km above the Earth surface. 
For solar neutrinos, we calculate the oscillation probability following~\cite{Maltoni:2015kca, Xu_2023}, with the solar nucleon density from GS98 data~\cite{Grevesse:1998bj}.
Details of our numerical methods can be found in our previous work~\cite{Fang:2024syu}.

In Fig.~\ref{fig:osc_fraction-all}, we illustrate the oscillation probability ratios between scenarios with and without the fifth force effect across different types of neutrino experiments. The left panel shows the survival probability ratio for LBL experiments as a function of $E_\nu$ and $L$, with parameters $g_A^{\mu \mu} \neq 0, m_{A^{\prime}}=10^{-14} \mathrm{eV}$, and $g_A^{\mu \mu} g_V^N=-10^{-48}$. The central panel presents the oscillation ratio for ATM in terms of $E_\nu$ and $\cos \theta$, with $m_{A^{\prime}}=10^{-14} \mathrm{eV}$ and $g_A^{\mu \mu} g_V^N=-10^{-50}$. The right panel displays the survival probability ratio for Solar as a function of $E_\nu$ and $g_A^{\mu \mu } g_V^N$ at a production location of $r=0.05 R_{\odot}$ and $m_{A^{\prime}}=10^{-17} \mathrm{eV}>1 / R_{\odot}$, with $R_{\odot}$ is the radii of Sun.

In both the left and central panels of Fig.~\ref{fig:osc_fraction-all}, the ratios display asymptotic behavior at high neutrino energies. In this limit, the vacuum oscillation term becomes small compared to the MSW potential and the fifth force potential. For large values of the product $ g_A^{\mu \mu} g_V^N $, the full Hamiltonian contains large contributions in the $11$ and $22$ terms, which suppresses $\nu_\mu \to \nu_e$ and $\nu_\mu \to \nu_\tau$ oscillations, resulting in a nearly unchanged survival probability $P^{\rm 5th}_{\mu \mu}$. In contrast, without the fifth force, $\nu_\mu \to \nu_\tau$ oscillations remain significant, consistent with the SM. Thus, the asymptotic lines in the left and central panels correspond to the minimum probability for two-flavor
$\nu_\mu \to \nu_\tau$ oscillations in the SM scenario. 
For solar neutrinos in the right panel of Fig.~\ref{fig:osc_fraction-all}, the fifth force effect depends on both the coupling strength and neutrino energy. Stronger couplings and higher energies lead to more significant deviations from SM predictions. For reactor LBL and MBL experiments,
we provide probability ratio plots in the Appendix.

\section{Analysis} 

After numerically calculating the transition and survival probabilities, we analyze various neutrino oscillation data to constraint the fifth force couplings.

For ATM neutrinos, we consider the Icecube DeepCore~\cite{IceCube:2011ucd, IceCube-PINGU:2014okk, Ishihara:2019aao, IceCubeCollaboration:2023wtb} which provides the 1-D events distribution respect to the neutrino energy $E_\nu$, the cosine zenith angle $\cos\theta$, and the $L/E_{\nu}$ ratio~\cite{IceCubeCollaboration:2023wtb}. To analyze the sensitivity of atmospheric neutrino data to the fifth force, we apply the \( \chi^2 \) method~\cite{Fogli:2002pt, Brzeminski:2022rkf}, defined as:
\begin{align}
\chi^2(N, O) = 2 \sum_{\alpha} \left( N_{\alpha} - O_{\alpha} + O_{\alpha} \ln \frac{O_{\alpha}}{N_{\alpha}} \right),
\label{eq:SK_chis}
\end{align}
where $N$ and $O$ is the expected and observed number of events respectively, and the index $\alpha$ represents the bins indexes for the events distribution given by the experiment. To calculate the expected number of events in the 5th force model, we divide $E$ and $\cos\theta$ into small 2-D bins. For each bin, the ratio of events in the fifth force model ($N_{ij}$) to those in the SM ($N_{ij}^0$) is given by:
\begin{equation}
\frac{N_{ij}}{N_{ij}^0} = \frac{\sum_{a} \iint_{ij} P_{a \rightarrow \mu/\bar{\mu}} \Phi_{a} \, dE \, d\cos\theta}{\sum_{a} \iint_{ij} P^0_{a \rightarrow \mu/\bar{\mu}} \Phi_{a} \, dE \, d\cos\theta},
\label{eq:signal_ratio_E}
\end{equation}
where $i$ and $j$ represents the bin indices for $E$ and $\cos\theta$, \( P_{a \rightarrow \mu/\bar{\mu}} \) is the neutrino oscillation probability for the fifth force model,  \( \Phi_a(E_\nu, \cos\theta) \) is the ATM neutrino flux at its production location before the oscillation process provided by~\cite{Honda:2015fha}, and \( a \) sums over \( e,\mu \) neutrinos and anti-neutrinos. Using Eq.~\eqref{eq:signal_ratio_E}, we rescale the SM event distribution to obtain the event distribution under the fifth force model.  

For accelerator LBL neutrinos, we analyze data from T2K~\cite{T2K:2011qtm, T2K:2023mcm}, which uses Super-Kamiokande~\cite{Super-Kamiokande:2002weg} as its far detector, located 295 km from the source. The detector is a large water Cherenkov device with a fiducial volume of 22.5 ktons. The near detector complex (ND280) is located 280 m from the source. The neutrino beam, produced by a proton beam dump, peaks at $\sim$0.6 GeV, as shown as the black triangle point in Fig.\ref{fig:osc_fraction-all}. For the fifth force model, the expected number of events is rescaled using Eq.~\eqref{eq:signal_ratio_E}, with the oscillation probability now depending only on neutrino energy 
$E_\nu$ due to the fixed baseline. The $\chi^2$ statistic is given by:
\begin{equation}
\chi^2 = \sum_{i} \frac{\left( N_i - O_i \right)^2}{\left(\delta O_i\right)^2},
\label{eq:chi2-for-solar}
\end{equation}
where $\delta O_i$ is the uncertainty of the observed signals in the $i$-th bin.
Observed and expected $\nu_\mu$ and $\bar{\nu}_{\mu}$ event rates and the neutrino parameters $\theta_{23}$ and $\Delta m^2_{23}$ are taken from Ref.~\cite{T2K:2023mcm}. 

For solar neutrinos, we use the data from BOREXINO~\cite{BOREXINO:2018ohr}, SNO+SK~\cite{Super-Kamiokande:2023jbt}, and calculate the $\chi^{2}$ as
\begin{equation}
\chi^2 = \sum_{i} \frac{\left( P_{ee}(E_i) - P_{ee}^{\rm obs}(E_i) \right)^2}{\left( \delta P_{ee}^{\rm obs}(E_i) \right)^2},
\label{eq:chi2-for-solar}
\end{equation}
where $P_{ee}(E_i)$ is the theoretical survival probability of electron neutrinos at energy $E_i$, and $P_{ee}^{\rm obs}(E_i)$ and $\delta P_{ee}^{\rm obs}(E_i)$ are the observed survival probability and its uncertainty. Observed values from BOREXINO ~\cite{BOREXINO:2018ohr} include: $P_{ee}^{\rm obs}$ for $pp$ neutrinos at 0.267 MeV is 0.57 $\pm$ 0.09, for ${}^{7}\text{Be}$ neutrinos at 0.862 MeV is 0.53 $\pm$ 0.05, for $pep$ neutrinos at 1.44 MeV is 0.43 $\pm$ 0.11. For high-energy region ${}^{8}\text{B}$ neutrinos, data points include: at 8.1 MeV, 0.37 $\pm$ 0.08; at 7.4 MeV, 0.39 $\pm$ 0.09; and at 9.7 MeV, 0.35 $\pm$ 0.09. The SNO+SK ~\cite{Super-Kamiokande:2023jbt} analysis provides a survival probability at $10 \mathrm{MeV}\left( { }^8\text{B} \text{~neutrinos}\right)$ of $0.308 \pm 0.015$, with very small uncertainty. These data, particularly the SNO+SK results, dominate the sensitivity to $g_A^{e e} g_V^N$ and $g_A^{\mu \mu} g_V^N$ due to their relevance at high energies, shown in Fig.~\ref{fig:osc_fraction-all}.

For reactor experiments, we use data from KamLAND~\cite{KamLAND:2010fvi, KamLAND:2013rgu, reactorneutrinoweb, Mueller:2011nm} and Daya Bay~\cite{DayaBay:2015ayh, DayaBay:2016ggj, DayaBay:2018yms, DayaBay:2022orm}. Since the analysis are similar and their constraints are weaker than the solar, ATM and accelerator neutrino results, we postpone the details of analysis into the Appendix.

\section{Muon $g-2$} 

External fields generated by dark matter~\cite{Graham:2020kai, Janish:2020knz} or fifth forces~\cite{Agrawal:2022wjm, Davoudiasl:2022gdg, Ema:2023pac, Fang:2024syu} may couple to muon spins, providing a testable signature in storage ring experiments and offering a potential solution to the muon g-2 anomaly~\cite{Davoudiasl:2022gdg, Fang:2024syu}.

\begin{figure*}[]
\begin{centering}
\includegraphics[width=1.0\textwidth]{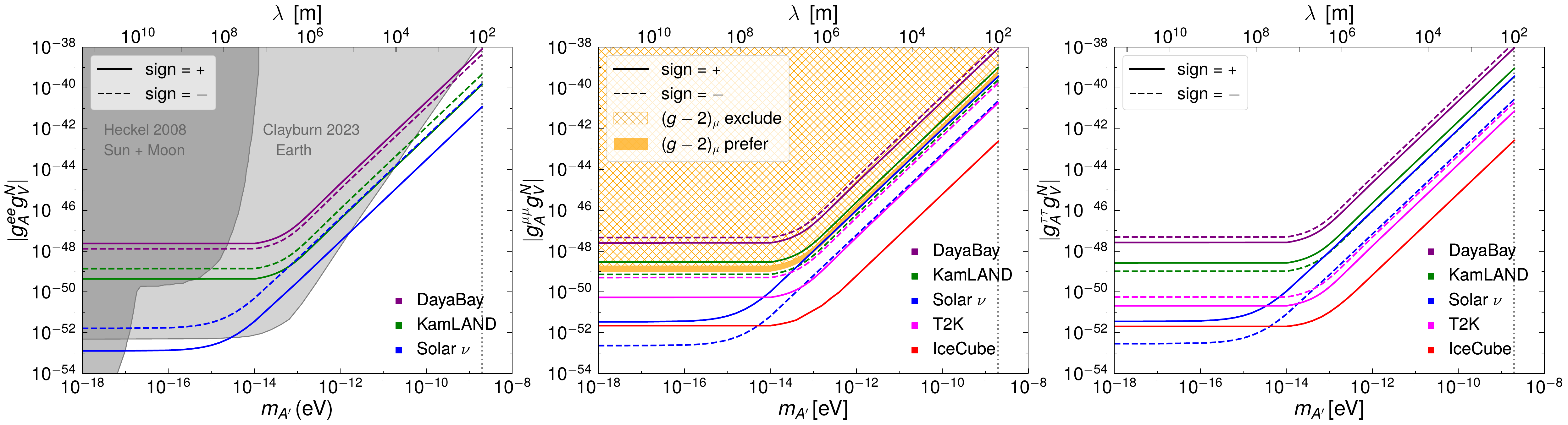}
\caption{
The 95\% C.L. constraints on the fifth force model with electron (left), muon (center) and tau (right) couplings. The lines show constraints from reactor experiments (Daya Bay~\cite{DayaBay:2015ayh, DayaBay:2016ggj, DayaBay:2018yms, DayaBay:2022orm}, KamLAND~\cite{KamLAND:2010fvi, KamLAND:2013rgu}), solar neutrino experiments (BOREXINO~\cite{BOREXINO:2018ohr}, SNO~\cite{SNO:2011hxd}, SK~\cite{Super-Kamiokande:2017yvm, Super-Kamiokande:2023ahc}), the accelerator experiment T2K~\cite{T2K:2011qtm, T2K:2023mcm}, and atmospheric neutrino data from IceCube DeepCore~\cite{IceCube:2011ucd, Ishihara:2019aao, IceCubeCollaboration:2023wtb}. Solid (dashed) lines represent positive (negative) values of the coupling constant \( g_A g_V^N \). The red dashed curve is omitted as the IceCube constraints for positive and negative couplings are nearly identical.
The existing constraints from spin-sensor experiments are shown in gray, labeled as Heckel 2008~\cite{Heckel:2008hw} and Clayburn 2023~\cite{Clayburn:2023eys}.
The orange hatched region (shaded region) shows parameter space excluded (preferred) by current muon \( g-2 \) measurements~\cite{Muong-2:2021ojo, Muong-2:2023cdq, Muong-2:2024hpx, ParticleDataGroup:2022pth}.
}
\label{fig:constraint} 
\end{centering}
\end{figure*}

If $A'$ has the $g_A^{\mu \mu}$ coupling, it contributes to $(g-2)_\mu$. In order to calculate its contribution, we boost the muon to its rotation muon rest frame (RMRF), where the muon is at rest~\cite{Graham:2020kai, Janish:2020knz}.
The fifth force Hamiltonian in RMRF is then given by~\cite{Dobrescu:2006au, Fadeev:2018rfl}: 
\be
H_{5{\rm th}} = g_A^{\mu\mu} \hat{\vec{\sigma}} \cdot \vec{\tilde{A}}' = -g_A^{\mu\mu} \hat{\vec{\sigma}} \cdot \frac{\vec{p}}{m_{\mu}} A'_0,
\label{Hami-g-2}
\ee
where $\vec{p}$ is the momentum of muon in the lab frame, $\tilde{A}'_\alpha \approx (\gamma A'_0, -\gamma \vec{\beta} A'_0) $ is the $A'$ field in RMRF, $\gamma \approx 29.3$ and $\vec{\beta}$ are the boost factor and velocity of the muon respectively. 
Using the spin evolution relations~\cite{Ema:2023pac, Fang:2024syu}, 
\be
\frac{d\vec{S}}{d t} = \vec{\omega} \times \vec{S} \,, \quad
\frac{d \hat{S}_i }{d t} = i\left[ H, \hat{S}_i \right],
\ee
where $ \hat{S}_i = \hat\sigma_i /2 $ is the muon spin operator.
We have the the precession frequency from the fifth force contribution:
\be
\delta \vec{\omega} = - 2 g_A^{\mu\mu} \frac{\vec{p}}{m_{\mu}} A'_0.
\label{eq:5th-deltaomegavec}
\ee
Since $\vec{p} \perp \vec \omega$, the precession frequency shift is~\cite{Graham:2020kai, Janish:2020knz}
\be
\Delta \omega = \sqrt{(\delta \vec\omega + \vec \omega)^2} - |\vec \omega|\approx |\delta \vec\omega|^2 / \left(2 |\vec \omega|\right).
\label{eq:5th-deltaomega}
\ee

Notably, $\Delta \omega$ is independent of the sign of the coupling constant $g_A^{\mu\mu} g_V^N$.
In RMRF, we have $|\vec \omega| = \gamma \omega_a \equiv \gamma (\omega_c -\omega_s)$
where $\omega_c$ ($\omega_s$) is the cyclotron (spin precession) frequency in the lab frame.
In RMRF, using Eq.~\eqref{eq:5th-deltaomegavec} and \eqref{eq:5th-deltaomega}, we have
\be
\Delta \omega =  \frac{2 g^2_A \left(\gamma^2 - 1\right) A'^2_0}{\gamma \omega_a}.
\ee
Back to the lab frame, $\Delta \omega_{\rm lab} = \Delta \omega/\gamma$. The relative frequency shift connects to the muon g-2 as $\Delta \omega_{\rm lab}/\omega_a \approx \Delta a_{\mu}/a_\mu$~\cite{Muong-2:2021vma, Muong-2:2021ojo, Davoudiasl:2022gdg}.
Given the recent result $\Delta a_{\mu} = (249 \pm 48) \times 10^{-11} > 0$~\cite{Muong-2:2021ojo, Muong-2:2023cdq,Muong-2:2024hpx, ParticleDataGroup:2022pth},
we find that, for $m_{A'} = 10^{-14}$ eV, 
the coupling range 
$|g^{\mu\mu}_A g_V^N| \in [7.2 \times 10^{-50}, 1.4 \times 10^{-49}]$
can fit the current muon g-2 anomaly at $3\sigma$. Coupling larger than $1.4 \times 10^{-49}$ are excluded.
Constraints for other $A'$ mass can be rescaled accordingly, with the results displayed as yellow grid regions in Fig.~\ref{fig:constraint}.

\section{Precision spin sensors} 

Laboratory experiments utilizing precision spin sensors can also probe $g_A^{ee} g_V^N $ couplings. In the non-relativistic limit, the Hamiltonian \eqref{Hami-g-2} can generate a spin-velocity dependent potential~\cite{Dobrescu:2006au, Fadeev:2018rfl}:
\be
V_{12+13} = g_A^{ee} g_V^N \vec{\sigma}_e \cdot \vec{v}_{\rm rel} \frac{1}{4 \pi r} e^{-m_{A'} r},
\label{eq:v1213}
\ee
where $\vec\sigma$ is the Pauli matrices for electrons, $\vec{v}_{\rm rel}$ and $r$ are the relative velocity and distance between electrons and nucleons.

For a very light mediator, the most relevant experiment is the torsion pendulum by Heckel et al.~\cite{Heckel:2008hw}. This setup uses a spin pendulum made of AlNiCo and ${\rm SmCo}_5$, materials with a significant number of polarized electrons. When an external field $\vec{\beta} $ couples to the spin, it contributes to the pendulum's energy as $E_p = - N_p \vec{\sigma}_p \cdot \vec{\beta} $, where $ N_p \sim 10^{23} $ is the net number of polarized spins, and $\vec{\sigma}_p $ represents the pendulum's spin orientation. This interaction produces a measurable torque $\vec{\tau} = N_p \vec{\sigma}_p \times \vec{\beta} $.
For a fifth force range $ \lambda \equiv m_{A'}^{-1} \gg 1.5 \times 10^{11} \, \mathrm{m} $ (Sun) and $ 4 \times 10^8 \, \mathrm{m} $ (Moon), limits are set at $ g_A^{ee} g_V^N $ values of $ (+0.2 \pm 1.2) \times 10^{-56} $ and $ (-3.1 \pm 2.4) \times 10^{-50} $, shown as solid gray lines in Fig.~\ref{fig:constraint}. 

Clayburn 2023~\cite{Clayburn:2023eys} considered Earth as a moving, unpolarized source, where particles within Earth have velocities different from electrons on the surface, thus, creating the potential $ V_{12+13} $. By analyzing various electron and nuclear spin sensors, they established bounds on orientation-dependent energy shifts. The Heckel torsion-pendulum result led to a new constraint of $ |g_A^{ee} g_V^N | < 10^{-52} $ for a force range $ \lambda > R_{\rm Earth} $, surpassing Heckel's old result~\cite{Heckel:2008hw} for $ \lambda \lesssim 10^{10} \, \mathrm{m} $. For smaller force ranges $ \lambda < R_{\rm Earth} $, the constraint is $\propto \lambda^{-3}$ from the volume integration of Eq.~\eqref{eq:v1213}. Note $v_{\rm rel}$ is proportional to $r$, thus canceling $r$ in the denominator. For $ \lambda < 100 \, \mathrm{m} $, density inhomogeneities near the detector limit experimental accuracy, which is the vertical gray dotted line in Fig.~\ref{fig:constraint}.

\section{Results}

We display constraints from various neutrino experiments using the $\chi^2$ analysis in Fig.~\ref{fig:constraint}. With two free parameters, \( m_{A'} \) and \( g_A g_V^N \), the 95\% confidence level corresponds to \( \Delta \chi^2 = 5.99 \). Due to potential inhomogeneities in Earth's local density, we consider only parameter space where \( \lambda \equiv 1/m_{A'} > 100 \) m~\cite{Clayburn:2023eys}.

For \( g_A^{ee} g_V^N \) couplings, the primary constraint arises from solar neutrino experiments, which limit \( g_A^{ee} g_V^N \) to \( [-16.5, 1.3] \times 10^{-53} \) in the massless limit and \( [-41.6, 3.0] \times 10^{-50} \times m_{A'}^2/(10^{-13} \, \mathrm{eV})^2 \) in the high-mass region. For positive \( g_A^{ee} g_V^N \), solar neutrino limits are stronger than those from Clayburn 2023 study~\cite{Clayburn:2023eys}. In the high-mass range, neutrino oscillation constraints scale as \( m_{A'}^{-2} \) (see Eq.~\eqref{eq:approx_potential_2}), while spin-sensor constraints scale as \( m_{A'}^{-3} \), making neutrino oscillations more sensitive for \( \lambda \lesssim 10^5 \, \mathrm{m} \). The reactor experiments KamLAND and Daya Bay provide constraints of \( [-13.9, 4.3] \times 10^{-50} \) and \( [-1.3, 2.3] \times 10^{-48} \) in the massless limit.

Solar neutrinos yield stronger constraints than reactor neutrinos due to the higher nucleon density and total mass of Sun and the higher energy of \( ^8 \)B neutrinos compared to reactor neutrinos. Additionally, solar constraints vary by an order of magnitude between positive and negative \( g_A^{ee} g_V^N \) values, as negative values fit solar data better than the SM~\cite{Super-Kamiokande:2022lyl}.

For \( g_A^{\mu \mu} g_V^N \) couplings, the strongest constraints come from ATM neutrino experiments using IceCube data, due to the high neutrino energy. These constraints are \( |g_A^{\mu\mu} g_V^N| < 1.2 \times 10^{-51} \times m_{A'}^2/(10^{-13} \, \mathrm{eV})^2 \) at high masses and \( |g_A^{\mu\mu} g_V^N| < 2.2 \times 10^{-52} \) in the massless limit. Solar neutrino and accelerator (T2K) experiments also impose strong constraints, with notable sign dependencies for \( g_A^{\mu \mu} g_V^N \), that are not observed in the IceCube results. Typically, if a positive \( g^{ee}_A g_V^N \) provides a stronger constraint than the negative, the reverse is true for \( g^{\mu \mu}_A g_V^N \) in the same experiment. Our analysis shows that solar, ATM, and accelerator neutrino experiments exclude the fifth force parameter space for \( g_A^{\mu \mu} g_V^N \) relevant to the muon \( g-2 \) anomaly. For \( g_A^{\tau \tau} g_V^N \) couplings, the results are similar to \( g_A^{\mu \mu} g_V^N \) couplings.

\section{Discussions}

We examined a fifth force model with lepton spin couplings mediated by a vector boson, using neutrino oscillation experiments as a probe. Our results reveal that neutrino experiments provide constraints on electron-nucleon couplings comparable to those from precision spin sensors, outperforming them in the high-mass region. For muon-nucleon couplings, neutrino oscillations yield constraints two orders of magnitude stronger than those from muon $ g-2 $ measurements. Tau-nucleon couplings are constrained by neutrino oscillations at a similar level to muon couplings. 
Our work highlights the capability of neutrino experiments to probe fifth force interactions across all three lepton generations.

\section*{Acknowledgement}

The work of J.L. is supported by the National Science Foundation of China under Grants No. 12475103,  12235001 and 12075005. 
The work of X.P.W. is supported by the National Science Foundation of China under Grants No. 12375095 and the Fundamental Research Funds for the Central Universities.

\section*{Appendix A: The Oscillation Probability for Reactor neutrino experiments}
\label{App-1}

We first present the numerical results for the oscillation probability ratios between the fifth force model and the SM for both LBL and MBL reactor experiments, derived from the Lagrangian discussed in the main text. Interestingly, our calculations reveal an asymptotic behavior in the oscillation probabilities for the LBL reactor case.

\begin{figure}[]
\begin{centering}
\includegraphics[width=0.4\textwidth]{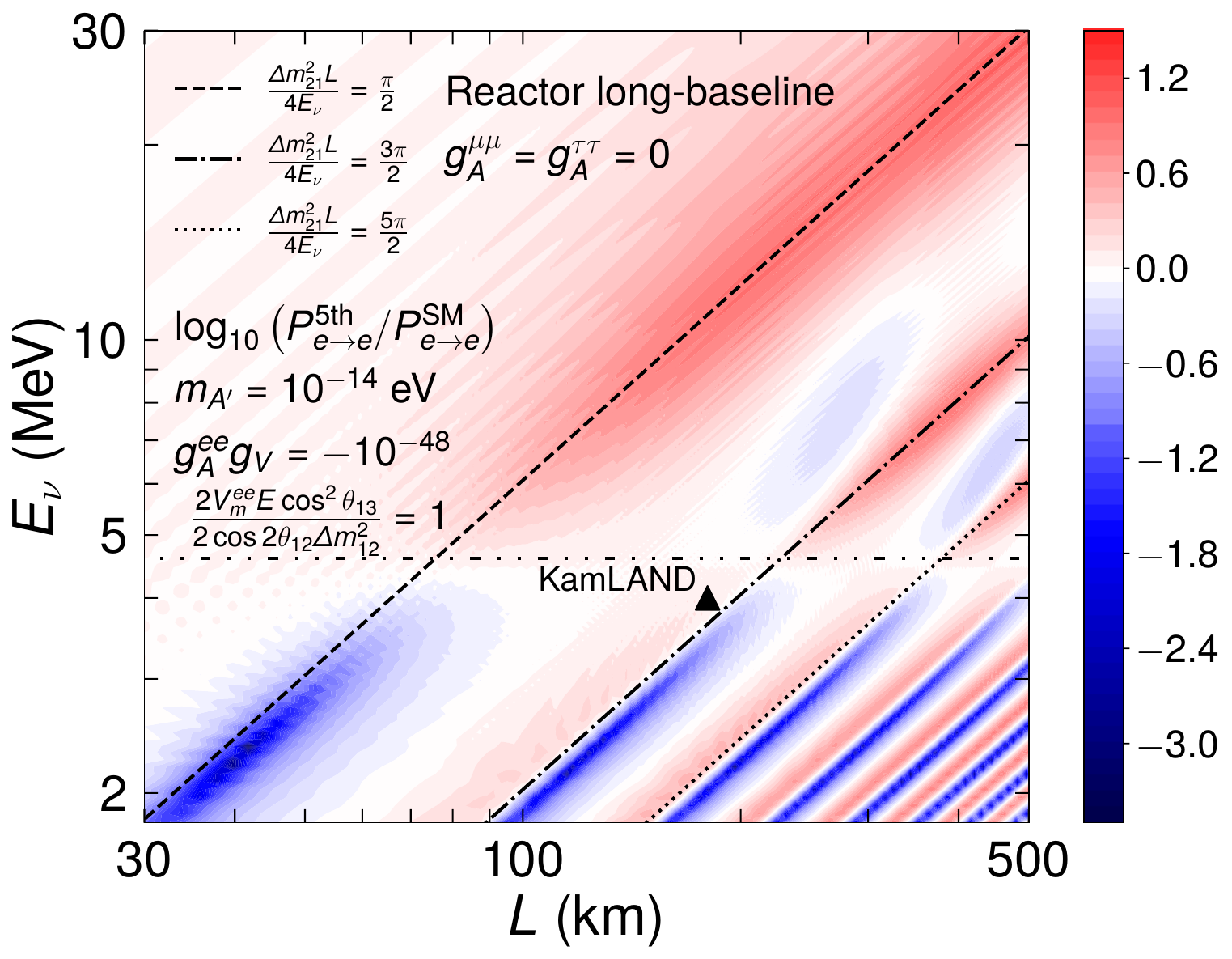}
\includegraphics[width=0.4\textwidth]{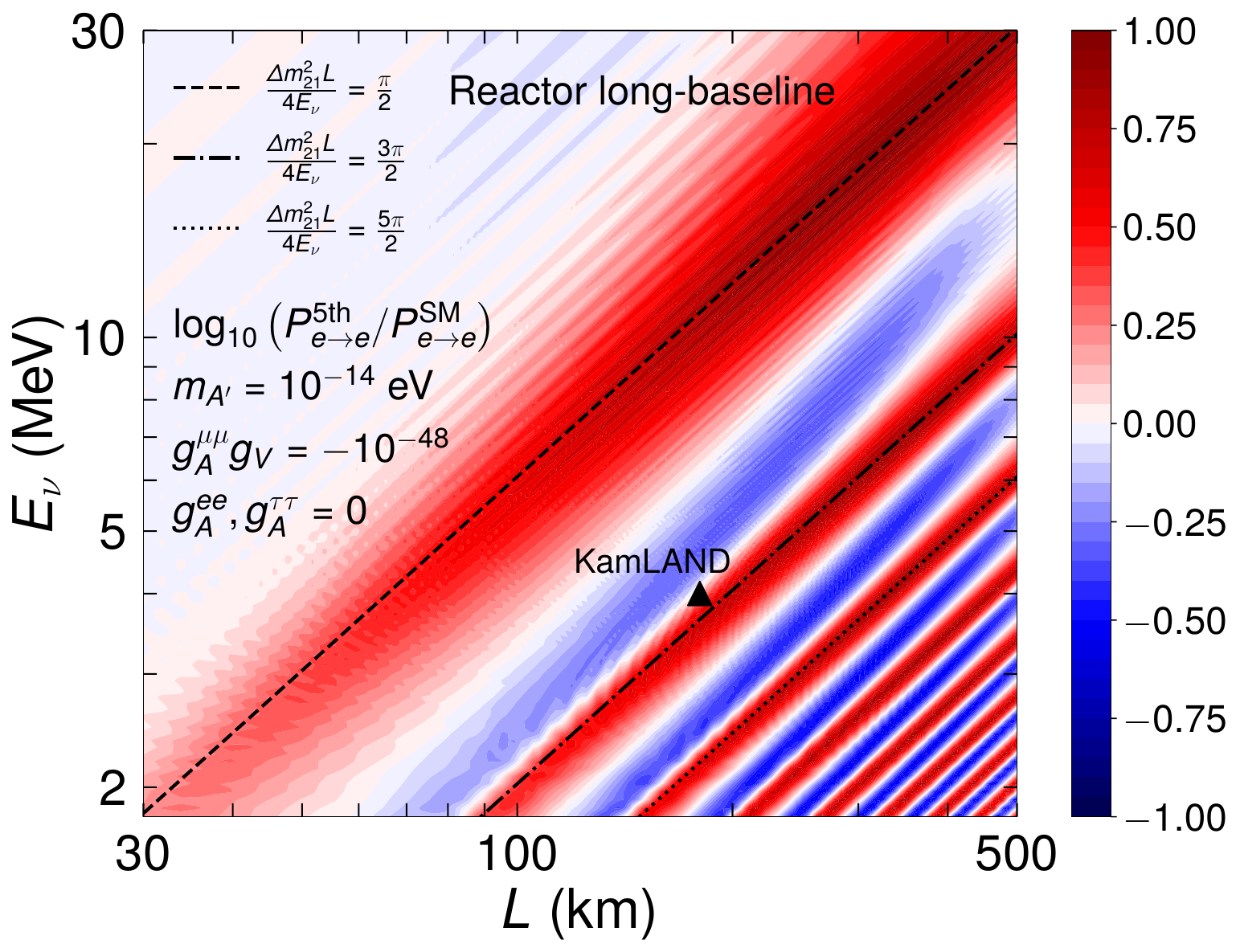}
\caption{Oscillation probability ratio between fifth force model and the SM for reactor LBL experiments.
}
\label{fig:osc_fraction_reactor_LBL} 
\end{centering}
\end{figure}

To analyze the impact of the fifth force on reactor LBL experiments, we fix \( m_{A'} = 10^{-14} \, \text{eV} \) and set \( g^{ee}_A g^N_V \) (\( g^{\mu\mu}_A g^N_V \)) to \( -10^{-48} \), respectively. The corresponding probability ratio plots are shown in Fig.~\ref{fig:osc_fraction_reactor_LBL}. In both cases, we observe that for high-energy neutrinos, the ratio reaches maximum deviations when 
\begin{align}
\frac{\Delta m^2_{21} L}{4 E_{\nu}} = \frac{(2N - 1) \pi}{2}, \quad N = 1, 2, 3, \ldots,
\end{align}
where $\Delta m^2_{21} L/4 E_{\nu} = \pi/2, \, 3\pi/2, \text{and } 5\pi/2$ are represented as dashed, dot-dashed, and dotted lines, respectively, in both plots of Fig.~\ref{fig:osc_fraction_reactor_LBL}. These asymptotic lines align closely with the regions where the probability ratios are largest. This behavior can explained that for reactor neutrinos, the MSW effects in the SM are negligible, and the oscillation driven by the \( \Delta m^2_{21} \) term dominates for baselines of approximately \( \mathcal{O}(100) \, \text{km} \). In contrast, under the fifth force model, the $ee$ terms in the full Hamiltonian become significantly larger than other terms for high-energy neutrinos. This suppresses the effective mixing angle between different flavors, driving the oscillation probability \( P^{5 \text{th}}_{e \to e} \) close to 1.

For the $g^{ee}_{A'}\neq 0$ case, the plot exhibits asymmetry around $2 V_m^{ee} E_\nu \cos^2 \theta_{13}/(2 \cos 2 \theta_{12} \Delta m_{12}^2) = 1 $, represented by the horizontal black dot-dashed line. As the energy decreases, the effective mixing angle and mass-squared difference in the fifth force model initially approach the corresponding values in the SM. Specifically, when~\cite{KamLAND:2013rgu}  
\begin{align}
      E_{\nu}= \frac{\Delta m^2_ {12} \cos 2 \theta_{12} }{  V^{ee}_m \cos^2 \theta_{13}},
\end{align}
the oscillation probabilities of the two models converge, as illustrated by the horizontal dot-dashed line in the upper panel of Fig.~\ref{fig:osc_fraction_reactor_LBL}. For even lower energies, the effective mixing angle in the fifth force model becomes larger than that in the SM, while the effective mass-squared difference becomes smaller~\cite{KamLAND:2013rgu}. As a result, in this energy regime, the probability ratio depends on oscillations from both the fifth force model and the SM.

\begin{figure}[htbp]
\begin{centering}
\includegraphics[width=0.4\textwidth]{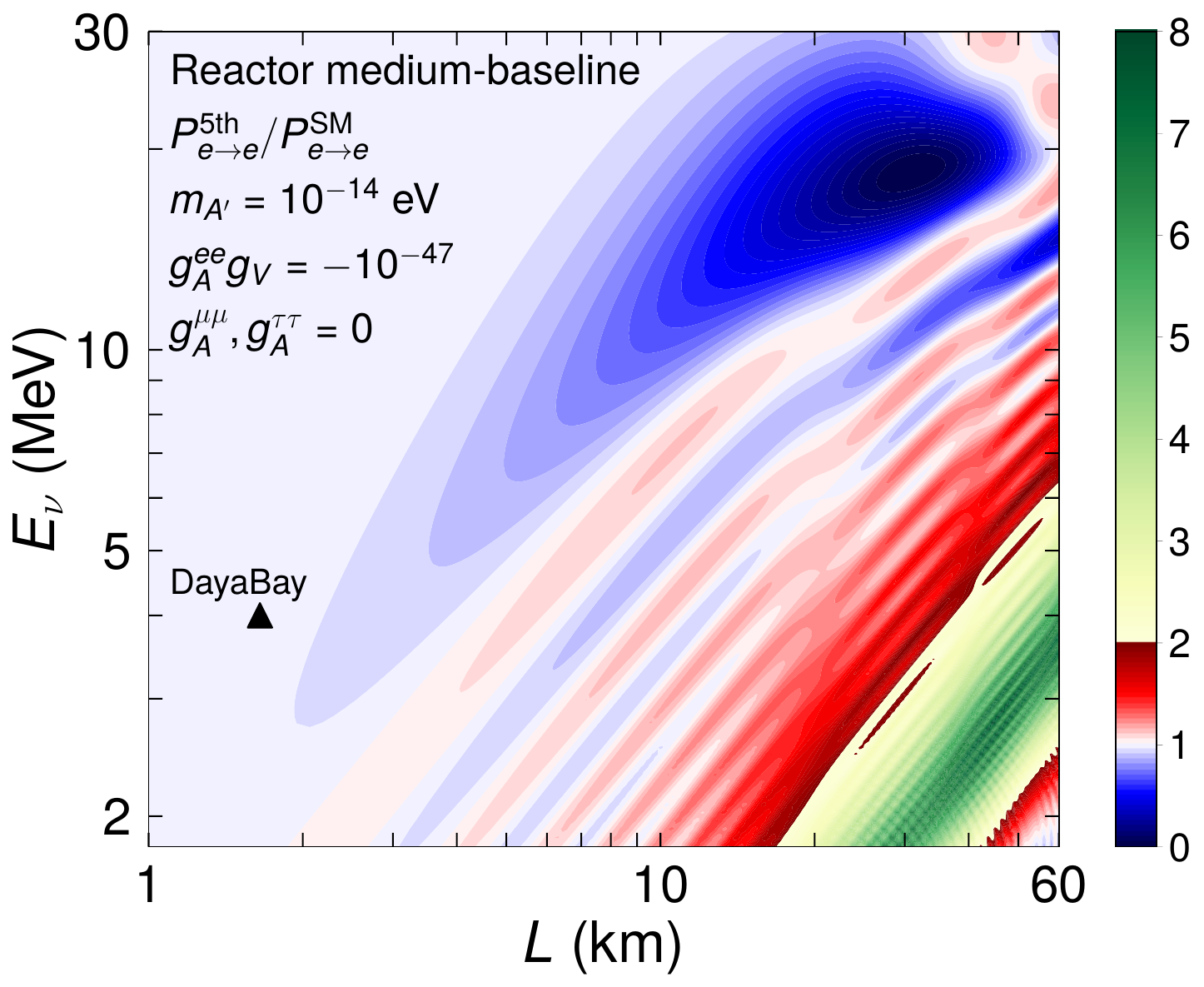}
\includegraphics[width=0.4\textwidth]{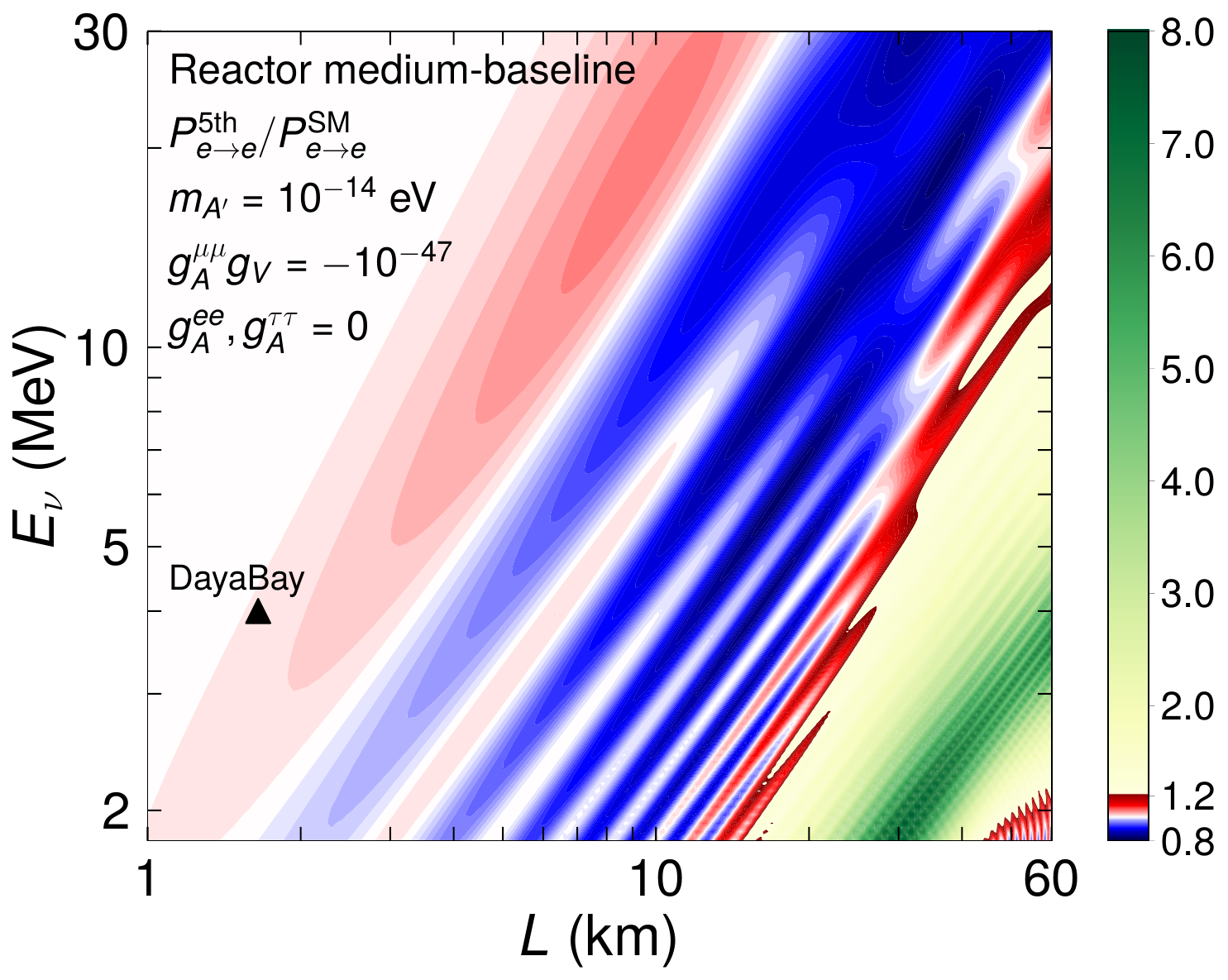}
\caption{Oscillation probability ratio between fifth force model and the SM for reactor MBL experiments.
}
\label{fig:osc_fraction_reactor_MBL} 
\end{centering}
\end{figure}

To evaluate the impact of the fifth force model on reactor MBL experiments, we set $m_{A'} = 10^{-14}$ eV and $g^{ee}_A g^N_V (g^{\mu\mu}_A g^N_V) = -10^{-47}$. Given the weaker oscillation effects at these distances, a larger coupling constant is used compared to the LBL cases. The resulting probability ratios are shown in Fig.~\ref{fig:osc_fraction_reactor_MBL}. For both cases, the ratio is significantly larger in the green-shaded region, as oscillations in this region are primarily driven by $\Delta m^2_{21}$, whose amplitude is much greater than that of oscillations dominated by $\Delta m^2_{31}$. Future experiments, such as JUNO, could potentially detect these effects in this energy and baseline range.

In Fig.~\ref{fig:osc_fraction_reactor_MBL}, for the case where only $g^{ee}_A \ne 0$, the ratio depends on both the vacuum oscillations in the SM and the oscillations induced by the fifth force model. Consequently, numerical methods are required to accurately evaluate its effects. Our analysis shows that the minimum ratio occurs when the neutrino energy is approximately 20 MeV and the baseline is around 30 km. For reactor neutrinos with an average energy of about 4 MeV, the minimum ratio is found near a 3 km baseline. This suggests that an experiment with a slightly longer baseline than Daya Bay could improve sensitivity to the fifth force model. In the case where $g^{\mu\mu}_A \ne 0$, vacuum oscillations play a more dominant role in shaping the ratios. The minimum ratio is found near points where ${\Delta m^2_{31} L}/{4 E_{\nu}} = 2N \pi /2$.

\section*{Appendix B: The analysis for Reactor neutrino experiments}
\label{App-2}
In this section, we detail the analysis procedure for reactor neutrino data and explain how the observed data can be used to derive the 95\% confidence level (C.L.) constraints on the fifth force model. 

\subsection{Reactor Long-baseline experiments}

For reactor LBL experiments, we use KamLAND~\cite{KamLAND:2010fvi, KamLAND:2013rgu} as an example. 
KamLAND consists of an inner detector containing approximately 1 kton of liquid scintillator as the target material and a 3.2 kton outer water-Cherenkov detector serving as a cosmic muon veto.
The neutrino flux at KamLAND
mainly comes from the operating reactors in Japan and Korean.
The flux-weighted average distance is about 180 km, marked as the black triangle in Fig.~\ref{fig:osc_fraction_reactor_LBL}.
This configuration makes KamLAND a critical experiment for precisely measuring the oscillation parameter $\Delta m^2_{12}$ and also
has important contributions in measuring $\theta_{12}$ and $\theta_{13}$~\cite{KamLAND:2013rgu}.

In our calculation, we take
the distances and total power output of 
all operating reactors in 
Japan and Korea from 2002 to 2012, as detailed in \cite{reactorneutrinoweb}.
The survival probability for each reactor
was firstly calculated individually based on its distance to KamLAND and
then averaged weighted by the total power.
For the averaged survival probability in each energy bin, we weighted the probabilities by the reactor neutrino flux:
\be
\overline{P^i_{ee}} = \frac{\int_i \sum_{k} f_k \Phi_k(E) P_{ee}(E) d E }{\int_i \sum_{k} f_k \Phi_k(E) d E},
\ee
where 
$i$ is the index of the energy bin,
$k$ represent the fission isotopics 
$({}^{235}\!\rm{U}, \allowbreak
{}^{238}\!\rm{U},
\allowbreak
{}^{239}\!\rm{Pu},
\allowbreak
{}^{241}\!\rm{Pu})$,
$f_k$ is
the contribution of each fission isotopic, taken to be 
$(0.567, \allowbreak 0.078, \allowbreak 0.298, \allowbreak 0.057)$~\cite{KamLAND:2013rgu}. $\Phi_k$ is the
anti-neutrino flux from the fission of each isotopic, as given in Ref.~\cite{Mueller:2011nm}.
Our results show that the calculated oscillation probabilities agree well with those in Ref.~\cite{KamLAND:2013rgu}.
Based on this consistency, we use our calculated survival probabilities to place constraints on the fifth force model using Eq.~\eqref{eq:chi2-for-solar} in the main text. The observed survival probabilities and uncertainties are taken directly from Ref.\cite{KamLAND:2013rgu}.

\subsection{Reactor Medium-baseline experiments}
For the reactor MBL neutrino experiment, we focus on the Daya Bay experiment~\cite{DayaBay:2015ayh,DayaBay:2016ggj,DayaBay:2018yms,DayaBay:2022orm}, which is known for its high-precision measurements of the oscillation parameters $\Delta m_{32}^{2}$ and $\sin^2 2\theta_{13}$. 
The experiment utilizes two nuclear power plant (NPP) complexes, Daya Bay and Ling Ao, which together host a total of six reactors—two at Daya Bay and four for Ling Ao NPP. 
The detection system includes two sets of near antineutrino detectors (ADs) located at experimental halls (EH1 and EH2) and one set of far ADs (at experimental hall EH3). 
Each detector has 20 tons of gadolinium doped liquid scintillator for detecting antineutrinos via the inverse beta decay process.
This configuration enables Daya Bay to achieve more precision in measuring $\Delta m_{32}^{2}$ compared to accelerator and atmospheric neutrino experiments.

Going through three stages with different amount of ADs, Ref.~\cite{DayaBay:2022orm}
provides the observed survival probability 
$P(\bar{\nu}_{e} \rightarrow \bar{\nu}_{e})$
corresponding to $L_{\rm eff}/\langle E_{\bar{\nu}_e} \rangle$, where 
$\langle E_{\bar{\nu}_e} \rangle$ is the mean antineutrino energy, and
the effective baseline $L_{\rm eff}$ represents the effective baseline. The effective baselines for EH1, EH2, and EH3 are 500 m, 500 m, and 1650 m, respectively \cite{DayaBay:2022orm}. Using the survival probabilities 
$P(\bar{\nu}_{e} \rightarrow \bar{\nu}_{e})$ observed at different energy points by the detectors in EH1, EH2, and EH3, we can place constraints on the fifth force model, as described using Eq.~\eqref{eq:chi2-for-solar} in the main text.

\bibliography{ref.bib}{}
\bibliographystyle{utphys28mod}
\end{document}